\documentclass[
twocolumn,
aps,
pra,
reprint,
superscriptaddress,
amsmath,amssymb
]{revtex4-1}
\usepackage[pdftex]{graphicx} 
\usepackage[hidelinks]{hyperref} 
\hypersetup{
 setpagesize=false,
 bookmarksnumbered=true,%
 bookmarksopen=true,%
 colorlinks=true,%
 linkcolor=blue,
 citecolor=blue,
}

\usepackage{braket}
\usepackage{times,multirow,amsfonts,bm,xspace,pifont,mathtools}
\usepackage{soul}
\begin{document}
\newcommand{\rr}{{\bm r}}
\newcommand{\q}{{\bm q}}
\renewcommand{\k}{{\bm k}}
\newcommand*\wien    {\textsc{wien}2k\xspace}
\newcommand*\textred[1]{\textcolor{red}{#1}}
\newcommand*\textblue[1]{\textcolor{blue}{#1}}
\newcommand*\YY[1]{\textcolor{red}{#1}}
\newcommand*\JI[1]{\textcolor{red}{#1}}
\newcommand{\ki}[1]{{\color{red}\st{#1}}}

\title{Periodic Anderson model for magnetism and superconductivity in \texorpdfstring{UTe$_2$}{UTe2}} 

\author{Jun Ishizuka}
\affiliation{Department of Physics, Graduate School of Science, Kyoto University, Kyoto 606-8502, Japan}

\author{Youichi Yanase}
\affiliation{Department of Physics, Graduate School of Science, Kyoto University, Kyoto 606-8502, Japan}
\affiliation{%
  Institute for Molecular Science, Okazaki 444-8585, Japan
}%
\date{\today}

\begin{abstract}
We provide and analyze a periodic Anderson model for studying magnetism and superconductivity in UTe$_2$, a recently-discovered candidate for a topological spin-triplet superconductor. The 24-band tight-binding model reproduces the band structure obtained from a DFT$+U$ calculation consistent with an angle-resolved photoemission spectroscopy. The Coulomb interaction of $f$-electrons enhances Ising ferromagnetic fluctuation along the $a$-axis and stabilizes spin-triplet superconductivity of either $B_{3u}$ or $A_{u}$ symmetry. When effects of pressure are taken into account in hopping integrals, the magnetic fluctuation changes to an antiferromagnetic one, and accordingly spin-singlet superconductivity of $A_{g}$ symmetry is stabilized. Based on the results, we propose pressure-temperature and magnetic field-temperature phase diagrams revealing multiple superconducting phases as well as an antiferromagnetic phase. In particular, a mixed-parity superconducting state with spontaneous inversion symmetry breaking is predicted. 


\end{abstract}

\maketitle

\section{Introduction}

A recent discovery of superconductivity in UTe$_2$~\cite{Ran_UTe2_2019} provides a new platform of spin-triplet superconductivity, which has been attracting renewed interest stimulated by the topological nature and accompanied Majorana fermion \cite{Qi_review,Sato_review2016, Sato_review2017}. Indeed, identifying the spin-triplet pairing state and pairing mechanism is one of the central topics in modern condensed matter physics.
Evidence for spin-triplet superconductivity in UTe$_2$ is manifested by an extremely large upper critical field \cite{Ran_UTe2_2019, Aoki_UTe2}, ferromagnetic fluctuation \cite{Tokunaga_UTe2, Sundar_UTe2}, reentrant superconductivity near metamagnetic transition \cite{Knebel_UTe2, Knafo_UTe2, A.Miyake_UTe2, Imajo_UTe2, Mineev_UTe2}, and NMR Knight shift revealing almost temperature-independent spin susceptibility below $T_{c}$ \cite{Nakamine_UTe2}. Existence of topological surface states expected in odd-parity superconductors was indeed reported~\cite{Jiao_UTe2_2020, Bae_UTe2}. Spin-triplet superconductivity is also implied by low-energy excitations measured by specific heat \cite{Ran_UTe2_2019,Kittaka_UTe2}, thermal conductivity \cite{Metz_UTe2,Izawa_private}, and magnetic penetration depth \cite{Metz_UTe2}, all of which are consistent with nodal quasiparticles.

Despite extensive research, the symmetry of superconductivity in UTe$_2$ still remains unsolved.
Time-reversal symmetry breaking has been reported by a scanning tunneling microscopy \cite{Jiao_UTe2_2020} and polar Kerr effect \cite{Hayes_UTe2_2020}, and a nonunitary chiral superconducting state 
has been proposed. However, the proposed chiral axes are different in the two studies.
Furthermore, the issue of whether the time-reversal symmetry breaking is an intrinsic property or not
needs to be solved by future studies.

Recent progress uncovered an impressive feature of this material, namely, multiple superconducting phases under pressure \cite{Braithwaite_UTe2_2019, Ran_UTe2_pressure, Lin_UTe2_2020, Knebel_UTe2_2020, Aoki_UTe2_2020, Thomas_UTe2_2020}.
A superconducting transition temperature $T_{c1}\sim1.6$ K is monotonically suppressed by pressure, and another superconducting phase appears with $T_{c2}$ increasing up to $3$ K at $P=1.2$ GPa. When the pressure is further increased, superconductivity is suppressed, and a potentially magnetic ordered state appears. 
An implication for an antiferromagnetic state is reported \cite{Aoki_UTe2_2020, Thomas_UTe2_2020} although UTe$_2$ has been considered to be near the ferromagnetic critical point. 
Furthermore, magnetic fields induce rich multiple superconducting phases under pressure \cite{Lin_UTe2_2020, Aoki_UTe2_2020, Thomas_UTe2_2020} as well as at ambient pressure~\cite{Knebel_UTe2,Ran_UTe2_H-T,Ishizuka_UTe2_2019}. 

From these observations, UTe$_2$ is expected to be a superconducting analog of superfluid $^3$He~\cite{Leggett1975} with multicomponent order parameters. 
However, different from $^3$He and another multicomponent superconductor UPt$_3$~\cite{Sauls1994,Joynt_UPt3}, the orthorhombic crystal structure of UTe$_2$ prohibits degenerate order parameters with the same $T_c$~\cite{Sigrist-Ueda}. Thus, accidental degeneracy not ensured by symmetry is required, and then, phenomenological theories \cite{Nevidomskyy_UTe2, Machida_UTe2} implementing symmetry constraint are less useful. On the other hand, the presence of the multiple superconducting phases is expected to be closely related to the magnetic phases, and therefore, theoretical studies linking superconductivity with magnetism are desired. For this purpose, a microscopic model for correlated electrons is needed. However, an effective Hamiltonian for UTe$_2$ has not been constructed. In order not only to clarify the pairing mechanism but also to identify the symmetry of multiple superconducting phases, the construction and analysis of a microscopic model for UTe$_2$ are highly awaited. Such a theory is also useful for uncovering topological superconducting phases because they can be specified by pairing symmetry, crystal structures, and Fermi surfaces (FSs)~\cite{Sato_review2016, Sato_review2017,Yanase2017,daido2019,Ono2019,Skurativska2020,Ono2020}.


A difficulty for theories of heavy fermion systems is a complex electronic structure. For this, first-principles calculations combined with experiments are informative. 
Density functional theory plus Hubbard $U$ (DFT$+U$) \cite{Shick_UTe2, Ishizuka_UTe2_2019, Xu_UTe2} and DFT combined with dynamical mean-field theory (DFT+DMFT) \cite{Xu_UTe2, Miao_UTe2} consistently predicted rectangular quasi-two-dimensional (2D) FSs for a large Coulomb interaction. Then, the FSs are formed by light electrons similar to ThTe$_2$ \cite{Miao_UTe2, Harima_UTe2}. On the other hand, for an intermediate Coulomb interaction $U$ 
we predicted an additional heavy FS around $\bm k = (0, 0, 2\pi)$~\cite{Ishizuka_UTe2_2019}, and it was indicated by angle-resolved photoemission spectroscopy (ARPES) \cite{Miao_UTe2}. 
This case realizes topological superconductivity 
\cite{Ishizuka_UTe2_2019}. 
Another ARPES study observed electron bands far below the Fermi level consistent with first-principles calculations~\cite{Fujimori_UTe2}.
A large carrier density is also compatible with thermoelectric power \cite{Niu_UTe2}. 



In this paper, we provide a minimal tight-binding model for UTe$_2$ based on the first-principles calculation for an intermediate $U$ and investigate magnetic fluctuation and superconductivity. 
Although we can successfully derive a realistic $72$-orbital model using the first-principles downfolding method, it is hard to study many-body effects in such a complicated model. Therefore, we here construct a $24$-band periodic Anderson model, which appropriately reproduces not only the topology of FSs but also the weight of U 5$f$, U 6$d$, and Te 5$p$ electrons obtained from DFT$+U$ calculations. 
The model predicts a reasonable pressure-temperature ($P$-$T$) phase diagram revealing spin-triplet superconductivity due to ferromagnetic fluctuation with the easy $a$-axis as well as spin-singlet superconductivity by antiferromagnetic fluctuations. 
From the result, 
we propose a mixed even-/odd-parity superconducting phase with spontaneous inversion symmetry breaking under pressure. 

\section{Model}
\begin{figure}[tbp]
\includegraphics[width=1.0\linewidth]{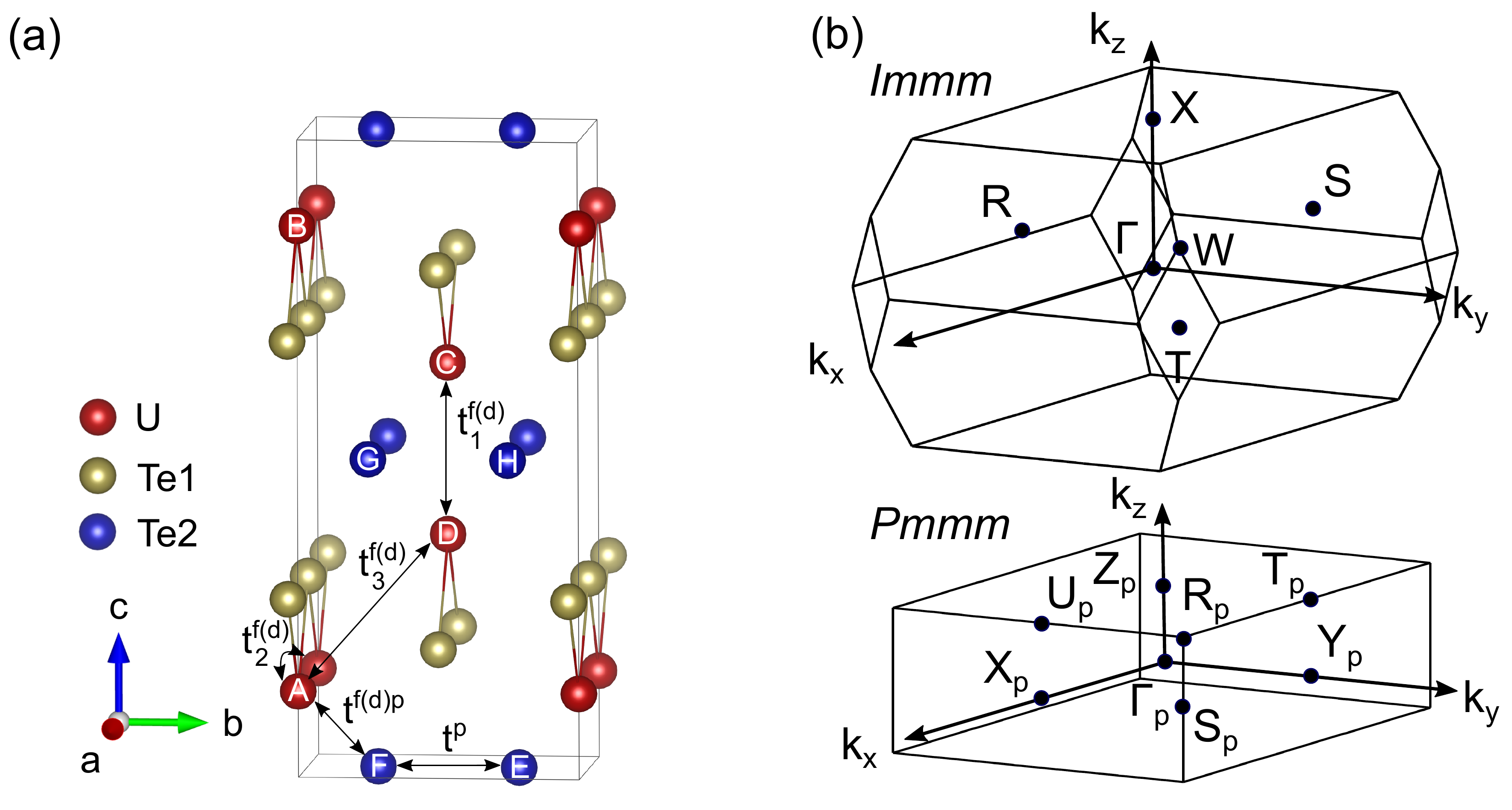}
\centering
\caption{
(a) Crystal structure and hopping integrals. (b) First BZ of original unit cell ($Immm$) and primitive unit cell ($Pmmm$).
\label{fig:StrBZ}}
\end{figure}

\begin{figure}[tbp]
\includegraphics[width=0.9\linewidth]{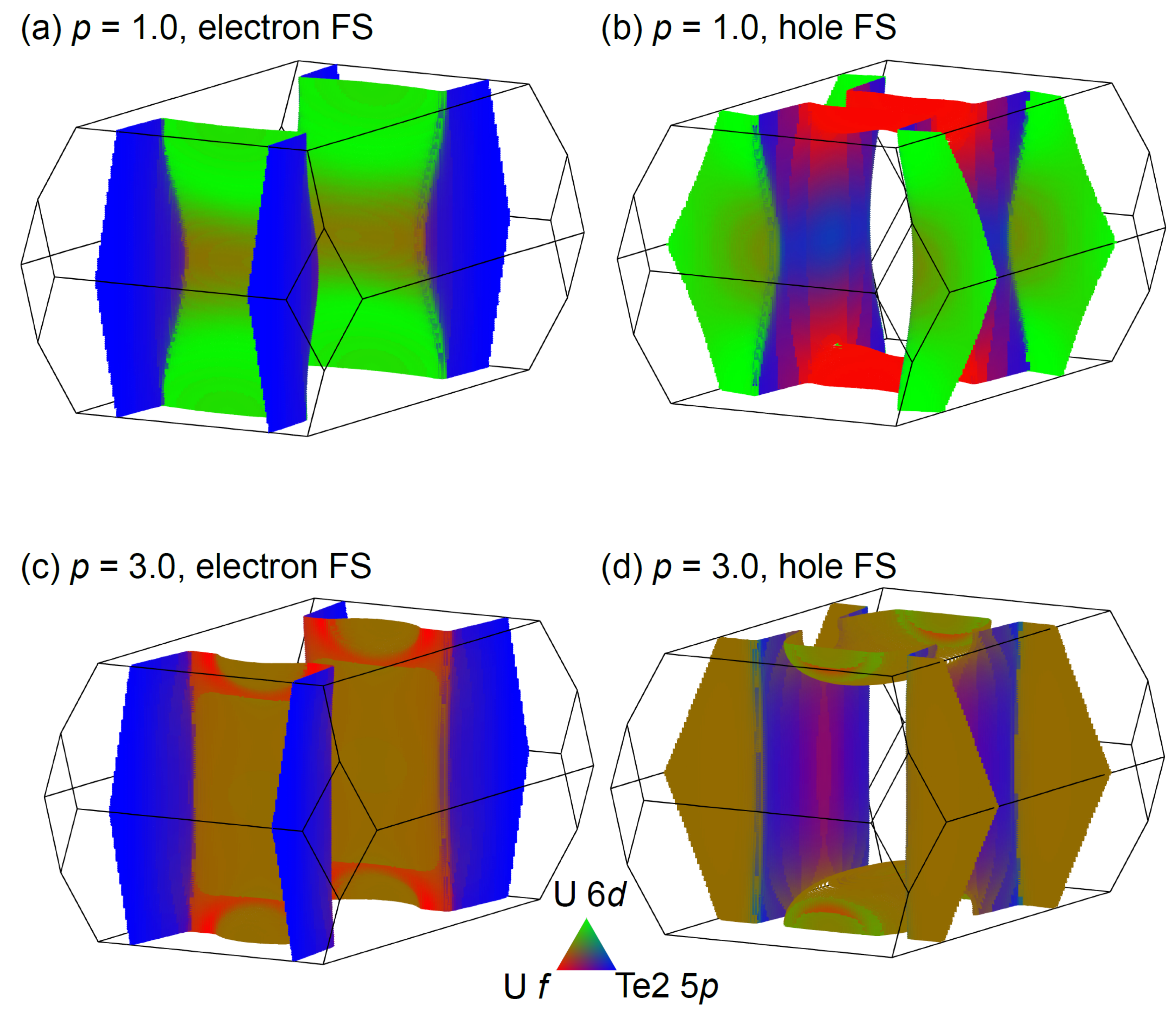} \centering
\caption{
Electron and hole FSs for (a)-(b) $p=1.0$ and (c)-(d) $p=3.0$. The blue, green, and red represent the weight of Te2 $5p$, U $6d$, and U $5f$ orbital, respectively. 
\label{fig:FS_I}}
\end{figure}

The DFT$+U$ calculations clarified hole and electron FSs, indicating a rather simple electronic structure near the Fermi level~\cite{Ishizuka_UTe2_2019,Xu_UTe2}. 
Therefore, we can construct a model including minimal hopping parameters [see Fig.~\ref{fig:StrBZ}(a)] which reproduces the low-energy electronic band structures in UTe$_2$. 
We adopt an original unit cell of the $Immm$ space group to illustrate FSs (Fig.~\ref{fig:FS_I}), while a primitive unit cell ($Pmmm$) with a folded Brillouin zone (BZ) [Fig.~\ref{fig:StrBZ}(b)] is adopted for convenience to calculate magnetic fluctuation and superconductivity. 
The resultant model is a $12$- or  $24$-band periodic Anderson model, when we take into account on-site Coulomb interaction of $f$-electrons. In addition, a sublattice-dependent antisymmetric spin-orbit coupling (sASOC) \cite{Sigrist_LNCS, Maruyama_LNCS, Fischer_noncentro, Yanase_UPt3_2016} is introduced in accordance with the local inversion symmetry breaking at uranium atoms.
Since uranium atoms form a ladder structure with local site symmetry $C_{2v}$, a Rashba-type sASOC appears with opposite coupling constants $\pm \alpha$. This sASOC induces magnetic anisotropy consistent with experiments~\cite{Ran_UTe2_2019,Aoki_UTe2}.
Here and hereafter, we set sASOC as $\alpha = 0.1$. 
Details of the tight-binding model are given in Appendix \ref{ap:tb}. 
We study the pressure effect by introducing an enhancement factor $p$ of hopping integrals. Hopping integrals of $f$-electrons and hybridization between $f$ and other orbitals are multiplied by $p \geq 1$, while $p=1$ at ambient pressure.
To translate $p$ into a real pressure, DFT calculations for UTe$_2$ under pressure are required. However, the lattice parameters under pressure have not been reported, and thus, we left it for a future study.

The band structure and FSs are shown in Figs.~S1 and \ref{fig:FS_I}. The weight of Te2 $5p$-, U $6d$-, and U $5f$-electrons is illustrated by color. 
The band structure exhibits flat $5f$-electron bands and one-dimensional dispersive $6d$- and $5p$-electron bands, each of which contributes to the FSs.
For a range of $1 \leq p \leq 3.5$, the topology of FS is consistent with ARPES \cite{Miao_UTe2} and DFT$+U$ calculations with intermediate $U$ \cite{Ishizuka_UTe2_2019}. 
Conducting directions of $5p$- and $6d$-electrons are orthogonal, and quasi-2D rectangular FSs are formed. Owing to the contribution of itinerant $f$-electrons the hole FS is bent and encloses $(0, 0, 2\pi)$ ($X$-point) as shown in Figs.~\ref{fig:FS_I}(b) and \ref{fig:FS_I}(d).
For $p=1.0$ [Fig.~\ref{fig:FS_I}(b)], a large $f$-orbital component is found near the $X$-point. 
When the factor $p$ is increased by pressure, the orbital character on FSs is largely changed, although FSs are only slightly changed. For $p=3.0$ [Fig.~\ref{fig:FS_I}(d)] we see a sizable $f$-electron component in a broad region on FSs. The change in orbital character results in a peculiar magnetic and superconducting phase diagram as we show below. 



\section{Magnetic fluctuation}
\begin{figure*}[tbp]
\includegraphics[width=1.0\linewidth]{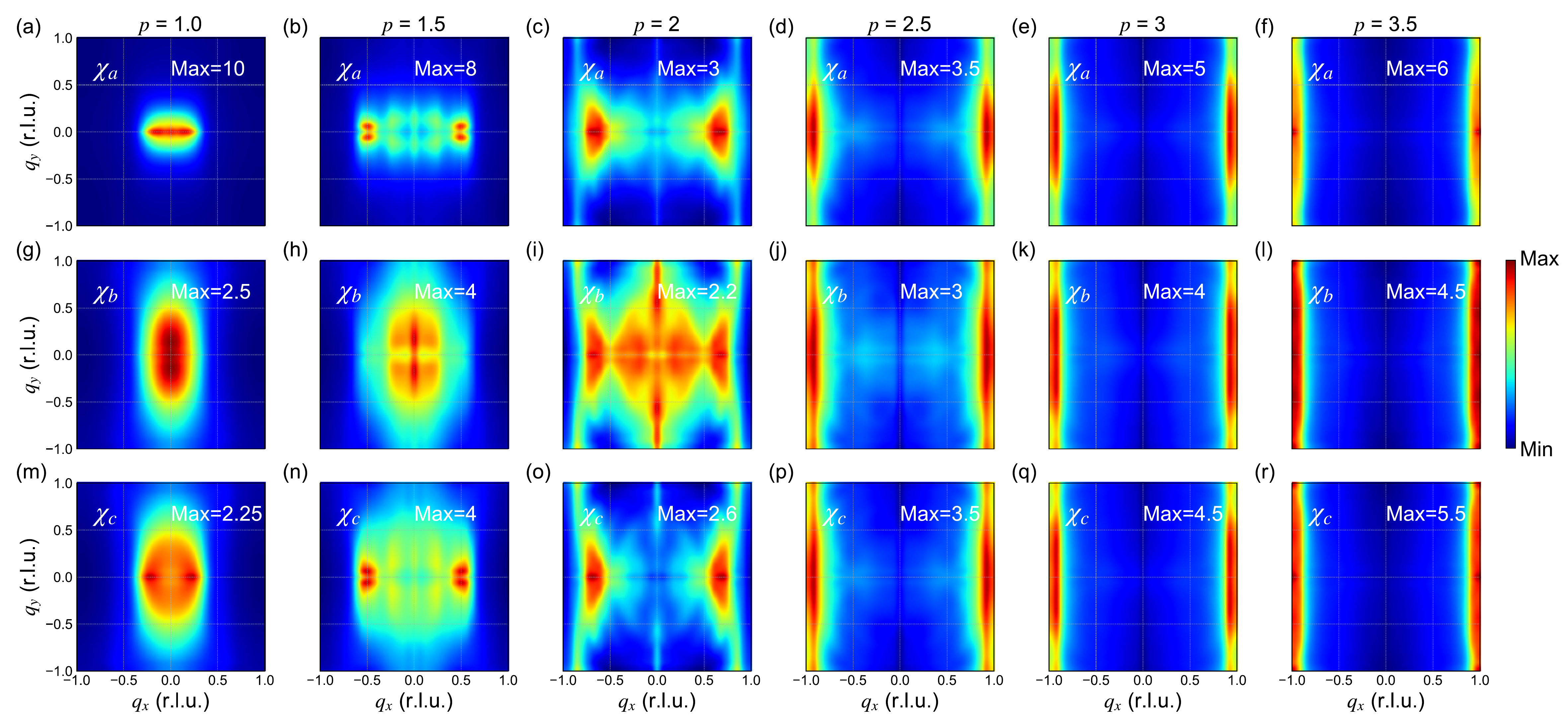}
\centering
\caption{
Magnetic susceptibility $\chi_a(\bm q, 0)$ (a-f), $\chi_b(\bm q, 0)$ (g-l), and $\chi_c(\bm q, 0)$ (m-r). Momentum dependence on the $q_x$-$q_y$ plane at $q_z = 0$ is drawn for $p=1.0$, $2.0$, and $3.0$ with $T=0.003$. We set $U=1.5$ for $p=1.0$ and $U=1.9$ for others.
\label{fig:chi}}
\end{figure*}

We apply the random phase approximation for the Coulomb interaction $U$ of $f$-electrons.
Diagonal magnetic susceptibilities, $\chi_a$, $\chi_b$, and $\chi_c$, are calculated from the susceptibility matrix of $f$-electrons (see Appendix \ref{ap:chi}), and the momentum dependence is shown in Fig. \ref{fig:chi}. 
For $p=1.0$, we see a ferromagnetic fluctuation with Ising anisotropy along the $a$-axis [Figs.~\ref{fig:chi}(a), \ref{fig:chi}(g), and \ref{fig:chi}(m)]; $\chi_a \simeq 10$ at maximum is much larger than $\chi_b \simeq 2.5$ and $\chi_c \simeq 2.25$.
This is in good agreement with experiments at ambient pressure~\cite{Ran_UTe2_2019,Aoki_UTe2}.
On the other hand, with increasing $p$, ferromagnetic fluctuation gradually changes to the antiferromagnetic fluctuation. This result implies that the magnetically ordered phase observed under pressure~\cite{Braithwaite_UTe2_2019, Knebel_UTe2_2020, Aoki_UTe2_2020, Thomas_UTe2_2020} is an antiferromagnetic phase.
The magnetic anisotropy at the ordering vector is reduced by pressure; for instance, $(\chi_a, \chi_b, \chi_c) \simeq (5, 4, 4.5)$ at ${\bm q} = (\pi,0,0)$ for $p = 3.0$. 

Growth of antiferromagnetic fluctuation originates from the change in orbital characters. Although the FSs show nesting property irrespective of the factor $p$, the $f$-electron component is negligible on the nested part of FSs for $p=1.0$. Therefore, $f$-electrons around the $X$-point enhance the ferromagnetic fluctuation rather than antiferromagnetic one. 
However, for $p=3.0$ the $f$-electron component is sizable on the nested FSs, and therefore, antiferromagnetic fluctuation develops around a nesting vector $\bm{q} = (\pi,0,0)$.
%
The $q$-vector corresponds to antiparallel alignment of the magnetic moment along Uranium chains. 
As for an intra-unit-cell structure, parallel alignment of magnetic moment between sublattices is favored.  
This means that, from the view point of augmented cluster multipole \cite{Watanabe_mpole, Hayami_Toroidal}, the obtained magnetic fluctuation is classified as even-parity magnetic dipole fluctuation, and the odd-parity magnetic fluctuation \cite{Ishizuka_Ir_2018} is not pronounced in UTe$_2$.


\section{Superconductivity}
Now, we clarify superconducting instability by solving the linearized Eliashberg equation (see Appendix \ref{ap:sc}).
In the $D_{2h}$ point group symmetry, the order parameter of superconductivity is classified as one of the eight irreducible representations.
In general, the Eliashberg equation is separable for each representation, and thus we obtain eight eigenvalues for each parameter set. Superconductivity occurs when the maximum eigenvalue is unity. Thus, we can determine what is the most stable superconducting state by comparing the eigenvalues.

\begin{figure}[tbp]
\includegraphics[width=0.9\linewidth]{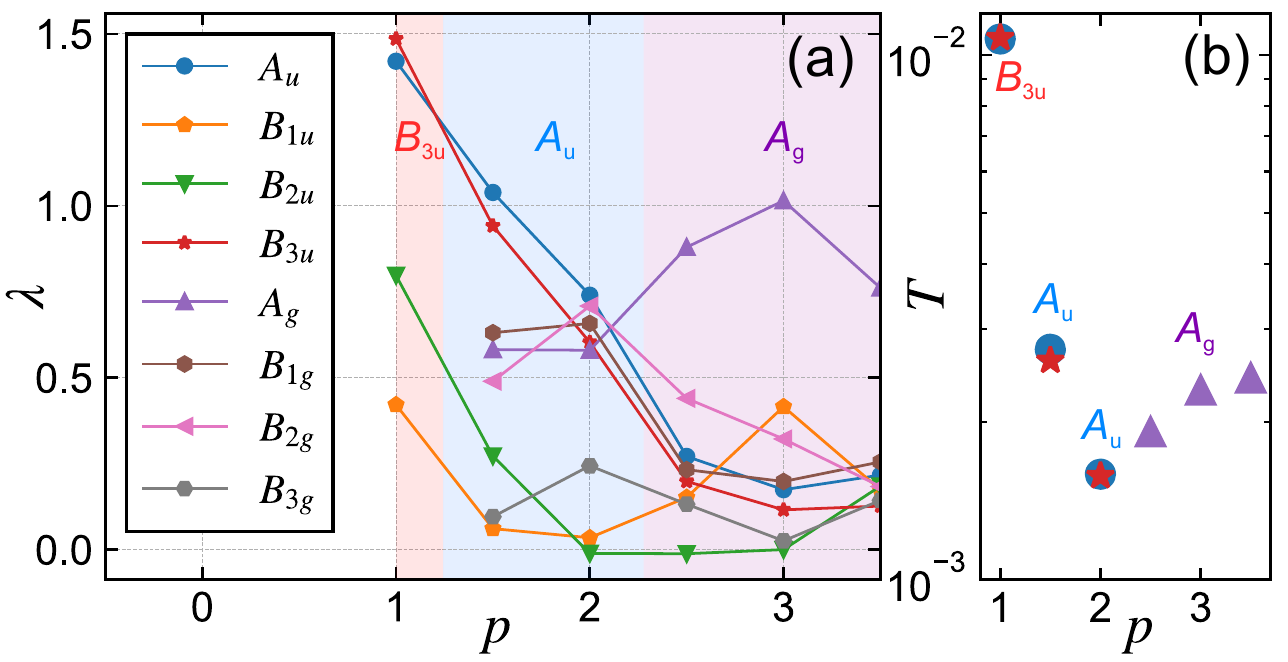}
\centering
\caption{
(a) Eigenvalues $\lambda$ of the Eliashberg equation for various irreducible representations of $D_{2h}$ point group. The parameter $p>1$ indicates applied pressure. We set $T=0.003$. The Coulomb interaction $U$ is set so that the Stoner factor is $\alpha_{\rm sf}=0.98$ and is shown in the Supplemental Material \cite{suppl}.
(b) Transition temperatures of the $A_u$, $B_{3u}$, and $A_g$ superconducting states 
with a fixed $U=1.9$. 
}
\label{fig:lambda}
\end{figure}

The $p$ dependence of eigenvalues is shown in Fig.~\ref{fig:lambda}(a). 
We see that the $B_{3u}$ pairing state is most stable at $p = 1.0$, indicating that the spin-triplet superconductivity is stabilized by the ferromagnetic fluctuation with Ising anisotropy.
For $p = 1.5$ and $2.0$, another spin-triplet pairing state with $A_u$ symmetry is stabilized by incommensurate magnetic fluctuations. The $B_{3u}$ and $A_u$ states are almost degenerate, and a more realistic model taking into account $j=5/2$ multiplet of $f$-electrons should be analyzed to compare the two states. 
Both $B_{3u}$ and $A_u$ states are monotonically suppressed by increasing $p$, and finally the $A_g$ state becomes predominant.
Thus, our results not only predict the odd-parity spin-triplet superconductivity of UTe$_2$ at ambient pressure but also indicate the spin-singlet superconductivity under pressure.
The latter is natural since the antiferromagnetic fluctuation usually stabilizes a $d$-wave or $s$-wave superconductivity, as widely believed for cuprates \cite{yanase2003} and iron-based superconductors \cite{Hosono_FeSC_2015}.

We also evaluate the critical temperature of superconducting instability based on the criterion $\lambda = 1$ and show the results in Fig.~\ref{fig:lambda}(b). 
The $B_{3u}$, $A_u$, and $A_g$ states may be stabilized below $T_{\rm c}$.
The pairing symmetry for each $p$ is consistent with Fig.~\ref{fig:lambda}(a). The transition temperature is highest at $p = 1$, in contrast to the experiment.
We need further study for quantifying the transition temperature and the pressure dependence, for instance, by conducting DFT calculations under pressure.

To clarify the $B_{3u}$, $A_u$, and $A_g$ states, we here discuss the order parameter of superconductivity.
In a standard manner, it is described as $\Delta(\boldsymbol{k}, i \pi T)=$ $\sum_{\mu} d^{\mu}(\bm k) \left[\sigma^{\mu} i \sigma^{y}\right]_{s s^{\prime}}$, with $\sigma^{\mu}$ the Pauli matrix for spin degree of freedom.
Although we omitted indices of sublattices for simplicity, intrasublattice and intersublattice components have similar structures (see Appendix \ref{ap:sc}). Thus, we show the maximum magnitude of intrasublattice components $d^{\mu}(\bm k)$ in whole momentum space (Table \ref{tab:gap}).
To be precise, all the states possess mixed spin-singlet and spin-triplet components since 
a sublattice-dependent parity mixing generally occurs in locally noncentrosymmetric systems~\cite{Maruyama_LNCS,Fischer_noncentro}. 
What kind of the parity mixing occurs is understood from the compatibility relation
, and the basis functions are given in Table \ref{tab:basis}.
According to Tables \ref{tab:gap} and \ref{tab:basis}, the predominant component for the $B_{3u}$ state is $d^{z}(\bm k) \simeq \gamma k_y $, 
while for the $A_u$ state it is $d^{y}(\bm k) \simeq \beta k_y$. 
Because these states are almost degenerate, the $d$-vector can rotate in the crystallographic $b$-$c$ plane. Thus, the Knight shift would be almost unchanged irrespective of the field direction, consistent with experimental results~\cite{Nakamine_UTe2,ishida_private}. 
For the $A_g$ state, a spin-singlet $s$-wave component with sign change, $d^{0}(\bm k) \simeq \delta \cos k_x$, is predominant. The subdominant spin-triplet component, $d^{y}(\bm k) \simeq \beta k_x$, is considerably small.

\begin{table}[tbp]
 \centering
 \caption{Maximum magnitudes of intrasublattice components of gap function $d^{\mu}(\bm k)$ obtained from the linearlized Eliashberg equation. The $B_{3u}$, $A_u$, and $A_{g}$ states for $p=1.0$, $2.0$, and $3.0$ are shown. Predominant components are labeled with a star $^\star$.}
 \label{tab:gap}
 \begin{tabular*}{1.0\columnwidth}{@{\extracolsep{\fill}}lcccc} \hline\hline
  & $d^{0}$ & $d^{x}$ & $d^{y}$ & $d^{z}$ \\ \hline
  $B_{3u}$ & $1.8 \times 10^{-5}$ & $2.5 \times 10^{-6}$ & $1.0 \times 10^{-4}$ & $^\star 4.3 \times 10^{-3}$ \\
  $A_{u}$ & $ 1.0 \times 10^{-3}$ & $ 4.0 \times 10^{-4}$ & $^\star 1.7 \times 10^{-3}$ & $ 3.8 \times 10^{-5}$ \\
  $A_{g}$ & $^\star 1.1 \times 10^{-3}$ & $2.5 \times 10^{-5}$ & $ 1.0 \times 10^{-4}$ & $ 7.4 \times 10^{-9}$  \\
 \hline\hline
 \end{tabular*}
\end{table}

\begin{table}[tbp]
 \centering
 \caption{Basis functions for the $B_{3u}$, $A_u$, and $A_g$ representations.}
 \label{tab:basis}
 \begin{tabular*}{1.0\columnwidth}{@{\extracolsep{\fill}}lcccc} \hline\hline
  & $d^{0}$ & $d^{x}$ & $d^{y}$ & $d^{z}$ \\ \hline
  $B_{3u}$ & $\delta k_x k_z$  & $\alpha k_x k_y k_z \hat{x}$ & $\beta k_z \hat{y}$ & $\gamma k_{y} \hat{z}$ \\
  $A_{u}$ & $\delta k_x k_y$ & $\alpha k_x \hat{x}$ & $\beta k_y \hat{y}$ & $\gamma k_{z} \hat{z}$ \\
  $A_{g}$ & $\delta k_x^2$ & $\alpha k_y \hat{x}$ & $\beta k_x \hat{y}$ & $\gamma k_x k_y k_z \hat{z}$  \\
 \hline\hline
 \end{tabular*}
\end{table}

\begin{figure*}[tbp]
\includegraphics[width=1.0\linewidth]{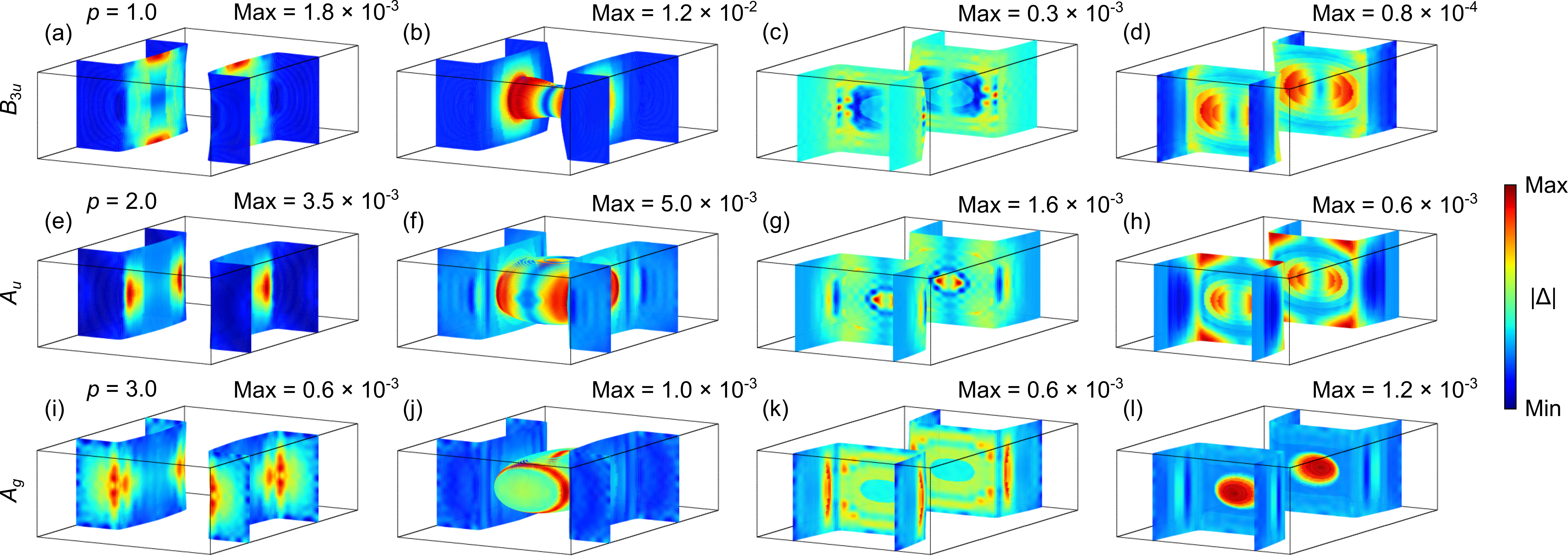}
\centering
\caption{
Superconducting gap structures on the electron and hole FSs obtained by the Eliashberg equation for $T=0.003$. The $B_{3u}$ (a-d), $A_u$ (e-h), and $A_g$ [(i-l)] states stabilized at $p=1.0$, $2.0$, and $3.0$ are shown. Since we adopt the primitive unit cell with folded BZ, the FSs are folded from those in the original BZ (Fig.~\ref{fig:FS_I}).
\label{fig:gap}}
\end{figure*}

Transforming to the band basis, we obtain superconducting gap structures illustrated in Fig.~\ref{fig:gap}.
As we considered the Coulomb interaction of $f$-electrons for superconductivity, the gap amplitudes are large on a part of FSs having a sizable contribution from $f$-electrons. Therefore, the gap structure is highly anisotropic in all the superconducting states. 
In accordance with group theories~\cite{Ishizuka_UTe2_2019, Yarzhemsky_UTe2}, we see symmetry-protected point nodes along the $k_x$ axis in the $B_{3u}$ state, while the line node is absent in agreement with Blount's theorem.
However, the gap minima, where the gap amplitude is not exactly zero, appears as a pseudoline node 
on the $k_y = 0$ plane [Fig.~\ref{fig:gap}(b)] because $|d^y(\bm k)| \ll |d^z(\bm k)|$. 
Similarly, the $A_u$ state shows a pseudo line node on $k_y = 0$, while it is a full-gap state in an exact sense.
It needs further investigations for the pseudo line node in comparison with experiments proposing point nodal gap \cite{Ran_UTe2_2019, Metz_UTe2} because the gap structure is significantly band dependent.

\section{Multiple superconducting phases}
\begin{figure}[tbp]
\includegraphics[width=1.0\linewidth]{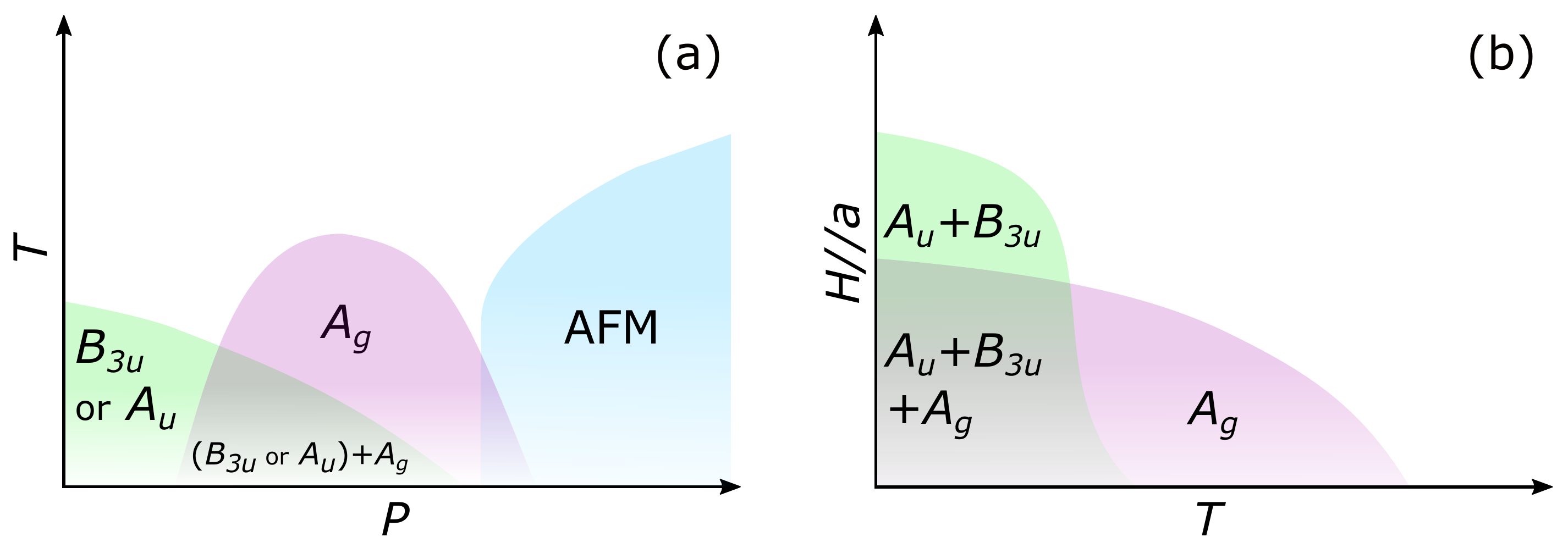}
\centering
\caption{
Proposed phase diagrams (a) in the $P$-$T$ plane, and (b) in the $T$-$H_a$ plane.
\label{fig:phase}}
\end{figure}

Based on the superconducting instability evaluated in Fig.~\ref{fig:lambda}(b), we illustrate our interpretation of the experimentally observed multiple superconducting phases in UTe$_2$~\cite{Braithwaite_UTe2_2019, Ran_UTe2_pressure, Lin_UTe2_2020, Knebel_UTe2_2020, Aoki_UTe2_2020, Thomas_UTe2_2020}. 
In Fig.~\ref{fig:phase}(a), we draw a superconducting phase transition from the odd-parity $B_{3u}$ or $A_u$ state to the even-parity $A_g$ state under the applied pressure, coinciding with crossover in magnetic fluctuations from ferromagnetic to antiferromagnetic. 
When the transition temperatures of the two states are close to each other, the coexistent phase is naturally expected, that is, either the $B_{3u} (A_{u}) + A_{g}$ or $B_{3u} (A_{u}) + i A_{g}$ state with mixed even-/odd-parity, and the space inversion symmetry is spontaneously broken. The time-reversal symmetry is preserved in the former, while the latter is $\mathcal{PT}$ symmetric. 

We also propose superconducting phases in the magnetic field $H \parallel a$ under pressure [Fig.~\ref{fig:phase}(b)]. In this magnetic field, $B_{3u}$ and $A_u$ representations are reduced to the same representation, and therefore, the $A_u + B_{3u}$ state is possible. This state almost avoids the paramagnetic depairing effect because the equal spin pairing along the $a$-axis is dominant. Thus, the upper critical field is naturally higher than that of the spin-singlet $A_{1g}$ state, and the superconducting phase diagram with a tricritical point is expected. 
Indeed, multiple superconducting phases as in Fig.~\ref{fig:phase}(b) have been reported in recent experiments~\cite{Knebel_UTe2_2020,Aoki_UTe2_2020}.

\section{Conclusion}
In this paper, we constructed a $24$-band periodic Anderson model as a reasonably realistic and easy-handled model for UTe$_2$. 
The model reveals not only the ferromagnetic fluctuation with the easy $a$-axis at ambient pressure but also the antiferromagnetic fluctuation under pressures. Accordingly, spin-triplet superconductivity of either $B_{3u}$ or $A_u$ representation is stabilized by the ferromagnetic fluctuation, while spin-singlet superconductivity of $A_g$ representation is favored by the antiferromagnetic fluctuation. 


These results enable us to draw phase diagrams in reasonable agreement with experiments. 
As a consequence, a mixed even-/odd-parity superconducting state with spontaneous inversion symmetry breaking is predicted. Such a phase was referred to in a review article~\cite{Leggett1975} forty years ago published with the comment "{\it there seems at present no experimental evidence.}"
Even at present, spontaneous ordering of mixed even-/odd-parity superconductivity has not been reported. UTe$_2$ may be the first material. Exploration of exotic superconducting properties will be the next issue. 

\begin{acknowledgments}
We appreciate helpful discussions with S. Kanasugi, K. Ishida, K. Izawa, S. Fujimori, D. Aoki, T. Shibauchi, A. H. Nevidomskyy, J. Flouquet, J.-P. Brison, and G. Knebel.
This work was supported by JSPS KAKENHI (Grants No.~JP18H04225, No.~JP18H05227, No.~JP18H01178, and No.~20H05159).
\end{acknowledgments}

\appendix
\section{Tight-binding model}
\label{ap:tb}
A periodic Anderson model is given by $H=H_{\rm t} + H_{\rm I}$, where $H_{\rm t}$ is the tight-binding model for a noninteracting part and $H_{\rm I}$ represents the on-site Coulomb interaction of $f$-electrons. 
Here, we introduce details of the tight-binding model for UTe$_2$,
\begin{align}
    H_{\rm t} = H_0 + H_{\rm ASOC}, \label{eq:H}
\end{align}
which contains a kinetic energy term $H_0$ and sublattice-dependent antisymmetric spin-orbit coupling (sASOC) term $H_{\rm ASOC}$.

The Hamiltonian of the kinetic energy term is given by
\begin{align}
    H_0 = \sum_{{\bm k},s} \hat a^\dag_{{\bm k} s} \begin{bmatrix}
        H_{\rm U}(\bm k) & H_{\rm U\mathchar`-Te}(\bm k) \\
        {\rm H.c.}       & H_{\rm Te}(\bm k)
    \end{bmatrix}
    \hat a_{{\bm k} s},
\end{align}
where
\begin{widetext}
\begin{align}
    H_{\rm U}(\bm k) =
    \begin{bmatrix}
    \varepsilon^f_{\rm AA}(\bm k) &     \varepsilon^{fd}_{\rm AA}(\bm k) & \varepsilon^f_{\rm AB}(\bm k) &
    \varepsilon^{fd}_{\rm AB}(\bm k) &
    &
    &
    \varepsilon^f_{\rm AD}(\bm k) &
    \varepsilon^{fd}_{\rm AD}(\bm k) & \\ 
    &
    \varepsilon^d_{\rm AA}(\bm k) &     \varepsilon^{fd}_{\rm AB}(\bm k) & \varepsilon^d_{\rm AB}(\bm k) &
    &
    &
    \varepsilon^{fd}_{\rm AD}(\bm k) &
    \varepsilon^{d}_{\rm AD}(\bm k) & \\ 
    &
    &
    \varepsilon^f_{\rm BB}(\bm k) &     \varepsilon^{fd}_{\rm BB}(\bm k) & \varepsilon^f_{\rm BC}(\bm k) &
    \varepsilon^{fd}_{\rm BC}(\bm k) &
    &
    & \\ 
    &
    &
    &
    \varepsilon^d_{\rm BB}(\bm k) &     \varepsilon^{fd}_{\rm BC}(\bm k) & \varepsilon^d_{\rm BC}(\bm k) &
    &
    & \\ 
    &
    &
    &
    &
    \varepsilon^f_{\rm CC}(\bm k) &     \varepsilon^{fd}_{\rm CC}(\bm k) & \varepsilon^f_{\rm CD}(\bm k) &
    \varepsilon^{fd}_{\rm CD}(\bm k) & \\ 
    &
    &
    &
    &
    &
    \varepsilon^d_{\rm CC}(\bm k) &     \varepsilon^{fd}_{\rm CD}(\bm k) & \varepsilon^d_{\rm CD}(\bm k) & \\ 
    &
    {\rm H.c.}&
    &
    &
    &
    &
    \varepsilon^f_{\rm DD}(\bm k) &     \varepsilon^{fd}_{\rm DD}(\bm k) & \\ 
    &
    &
    &
    &
    &
    &
    &
    \varepsilon^{d}_{\rm DD}(\bm k) & \\ 
    \end{bmatrix},
\end{align}
\end{widetext}
\begin{align}
    H_{\rm U\mathchar`-Te}(\bm k) =
    \begin{bmatrix}
    \varepsilon^{fp}_{\rm AE}(\bm k) &
    \varepsilon^{fp}_{\rm AF}(\bm k) & 
    &
    & \\ 
    \varepsilon^{dp}_{\rm AE}(\bm k) &
    \varepsilon^{dp}_{\rm AF}(\bm k) & 
    &
    & \\ 
    \varepsilon^{fp}_{\rm BE}(\bm k) &
    \varepsilon^{fp}_{\rm BF}(\bm k) & 
    &
    & \\ 
    \varepsilon^{dp}_{\rm BE}(\bm k) &
    \varepsilon^{dp}_{\rm BF}(\bm k) & 
    &
    & \\ 
    &
    &
    \varepsilon^{fp}_{\rm CG}(\bm k) &
    \varepsilon^{fp}_{\rm CH}(\bm k) & \\ 
    &
    &
    \varepsilon^{dp}_{\rm CG}(\bm k) &
    \varepsilon^{dp}_{\rm CH}(\bm k) & \\ 
    &
    &
    \varepsilon^{fp}_{\rm DG}(\bm k) &
    \varepsilon^{fp}_{\rm DH}(\bm k) & \\ 
    &
    &
    \varepsilon^{dp}_{\rm DG}(\bm k) &
    \varepsilon^{dp}_{\rm DH}(\bm k) & \\ 
    \end{bmatrix},
\end{align}
\begin{align}
    H_{\rm Te}(\bm k) =
    \begin{bmatrix}
    \varepsilon^p_{0} &
    \varepsilon^p_{\rm EF}(\bm k) &
    &
    & \\ 
    {\rm H.c.} &
    \varepsilon^p_{0} &
    &
    & \\ 
    &
    &
    \varepsilon^p_{0} &
    \varepsilon^p_{\rm GH}(\bm k) & \\ 
    &
    &
    {\rm H.c.} &
    \varepsilon^p_{0} & \\ 
    \end{bmatrix}, 
\end{align}
and 
\begin{widetext}
\begin{align}
\hat a^\dag_{{\bm k} s} = \left(
f^{\dag}_{\k {\rm A} s}, 
d^{\dag}_{\k {\rm A} s}, 
f^{\dag}_{\k {\rm B} s}, 
d^{\dag}_{\k {\rm B} s}, 
f^{\dag}_{\k {\rm C} s},
d^{\dag}_{\k {\rm C} s}, 
f^{\dag}_{\k {\rm D} s}, 
d^{\dag}_{\k {\rm D} s}, 
p^{\dag}_{\k {\rm E} s}, 
p^{\dag}_{\k {\rm F} s}, 
p^{\dag}_{\k {\rm G} s}, 
p^{\dag}_{\k {\rm H} s} \right)
\end{align}
\end{widetext}
with the primitive unit cell $\{a\hat x, b\hat y, c \hat z\}$. The annihilation (creation) operators of U $5f$, U $6d$, and Te2 $5p$ electrons with pseudospin $s$ on a sublattice $m=(\rm A, \rm B, \rm C, \rm D)$ and $\bar m=(\rm E, \rm F, \rm G, \rm H)$ are represented by $f^{(\dag)}_{\k m s}$, $d^{(\dag)}_{\k m s}$, and $p^{(\dag)}_{\k \bar m s}$, respectively.
The single-electron kinetic energy is described by taking into account the hopping integrals up to the third order shown in Fig.~\ref{fig:StrBZ} of the main text,
\begin{align}
    \varepsilon^{f(d)(fd)}_{\rm AA} (\k) & = \varepsilon^{f(d)(fd)}_{\rm BB} (\k) =
    \varepsilon^{f(d)(fd)}_{\rm CC} (\k) =
    \varepsilon^{f(d)(fd)}_{\rm DD} (\k) \nonumber \\
    & = \varepsilon^{f(d)}_0 + t^{f(d)(fd)}_2 (e^{ik_xa} + e^{ - ik_xa}), \\ 
    \varepsilon^{f(d)(fd)}_{\rm AB} (\k) & = t^{f(d)(fd)}_1 e^{ik_zc}, \\ 
    \varepsilon^{f(d)(fd)}_{\rm CD} (\k) & = t^{f(d)(fd)}_1, \\ 
    \varepsilon^{f(d)(fd)}_{\rm AD} (\k) & =
    \varepsilon^{f(d)(fd)}_{\rm BC} (\k) \nonumber \\
    & = t^{f(d)(fd)}_3 (1 + e^{ik_xa} + e^{ik_yb} + e^{ik_xa + ik_yb}), 
\end{align}
\begin{align}
    \varepsilon^{fp(dp)}_{\rm AE} (\k) & = t^{fp(dp)} (1 + e^{ik_xa}) e^{ik_yb}, \\ 
    \varepsilon^{fp(dp)}_{\rm AF} (\k) & = - t^{fp(dp)} (1 + e^{ik_xa}), \\ 
    \varepsilon^{fp(dp)}_{\rm BE} (\k) & = - t^{fp(dp)} (1 + e^{ik_xa}) e^{ik_yb} e^{ - ik_zc}, \\ 
    \varepsilon^{fp(dp)}_{\rm BF} (\k) & = t^{fp(dp)} (1 + e^{ik_xa}) e^{ - ik_zc}, \\ 
    \varepsilon^{fp(dp)}_{\rm CG} (\k) & = \varepsilon^{fp(dp)}_{\rm DH} (\k)
    = t^{fp(dp)} (1 + e^{ - ik_xa}), \\ 
    \varepsilon^{fp(dp)}_{\rm CH} (\k) & = \varepsilon^{fp(dp)}_{\rm DG} (\k)
    = - t^{fp(dp)} (1 + e^{ - ik_xa}), \\ 
    \varepsilon^{p}_{\rm EF} (\k) & = t^{p} (1 + e^{ - ik_yb}), \\ 
    \varepsilon^{p}_{\rm GH} (\k) & = t^{p} (1 + e^{ik_yb}). 
\end{align}
The tight-binding parameters in Table \ref{tab:params} reproduce the Fermi surfaces (FSs) observed in an ARPES experiment~\cite{Miao_UTe2} and DFT$+U$ calculations~\cite{Ishizuka_UTe2_2019}. As shown in the main text, the model for this parameter set shows enhanced ferromagnetic fluctuation with the easy $a$-axis in agreement with experiments at ambient pressure~\cite{Ran_UTe2_2019,Aoki_UTe2}. To investigate effects of pressure, we introduce an enhancement factor $p$ for the hopping integrals concerned with the $f$ electrons. The corresponding tight-binding parameters $\{t^f_1, t^f_2, t^f_3, t^{fd}_1, t^{fd}_2, t^{fd}_3, t^{fp}\}$ are multiplied by $p$ with $1<p<3.5$, while $p=1$ at ambient pressure.

The sASOC term is written as
\begin{align}
    H_{\mathrm{ASOC}} = & \left(\alpha_1 \sin k_y \hat{\sigma}_x - \alpha_2 \sin k_x \hat{\sigma}_y \right) \nonumber \\
    & \otimes \hat{\tau}^{\rm (intra)}_z \otimes \hat{\tau}^{\rm (inter)}_0,
\end{align}
where $\hat{\sigma}_i$, $\hat{\tau}^{\rm (intra)}_i$, and $\hat{\tau}^{\rm (inter)}_i$
are the Pauli matrices representing the spin, intra-ladder sublattice, and inter-ladder sublattice degrees of freedom, respectively.
We set $\alpha_1 = \alpha_2 = 0.1$ for simplicity.

\begin{figure}[tbp]
\includegraphics[width=1.0\linewidth]{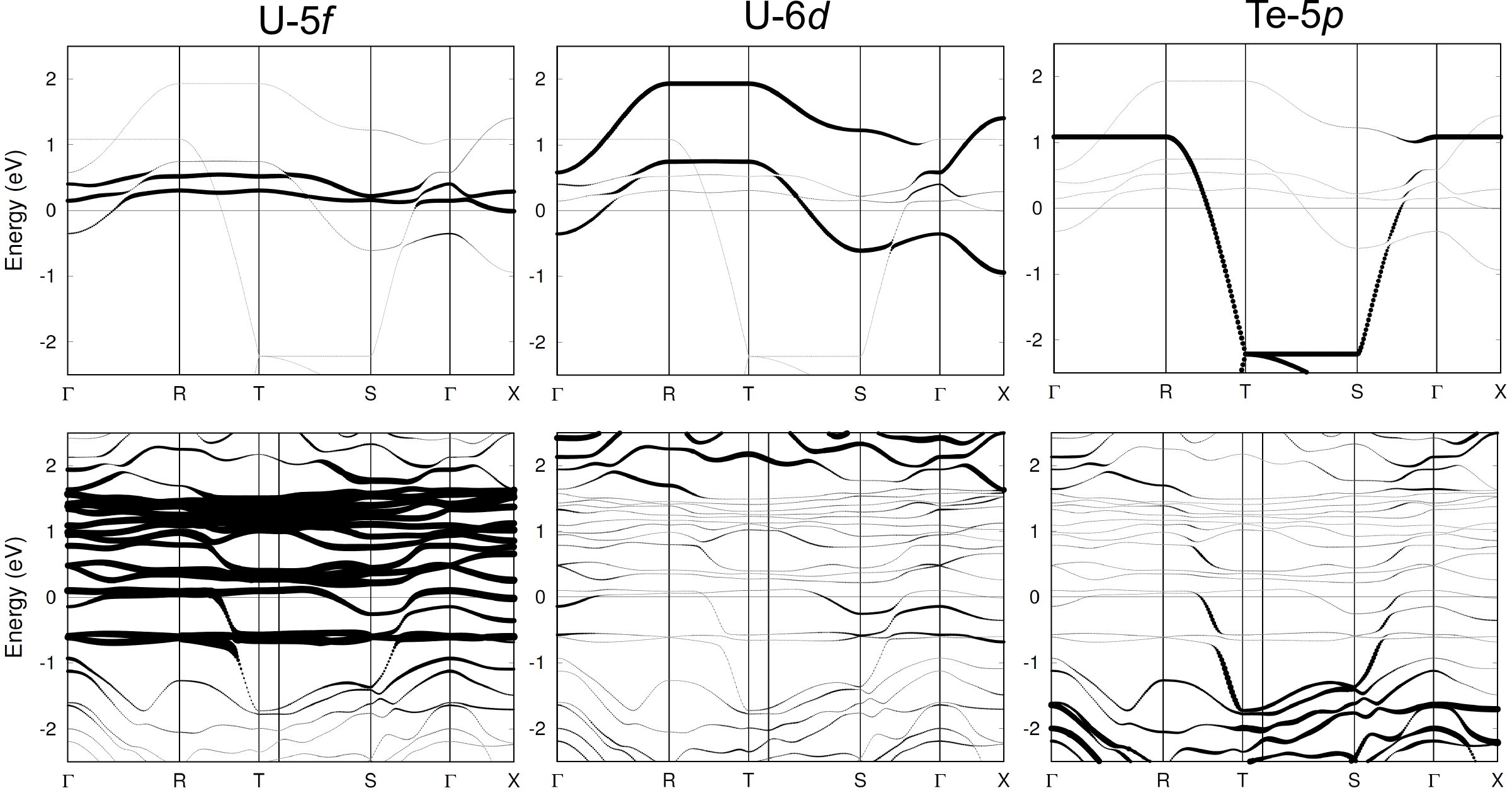}
\centering
\caption{
(Upper panel) Band structure of the $f$-$d$-$p$ tight-binding model described in Eq.~(\ref{eq:H}). (Lower panel) Band structure obtained from a DFT$+U$ calculation for $U=1.5$ eV.
\label{fig:band}}
\end{figure}

\begin{figure}[tbp]
\includegraphics[width=0.8\linewidth]{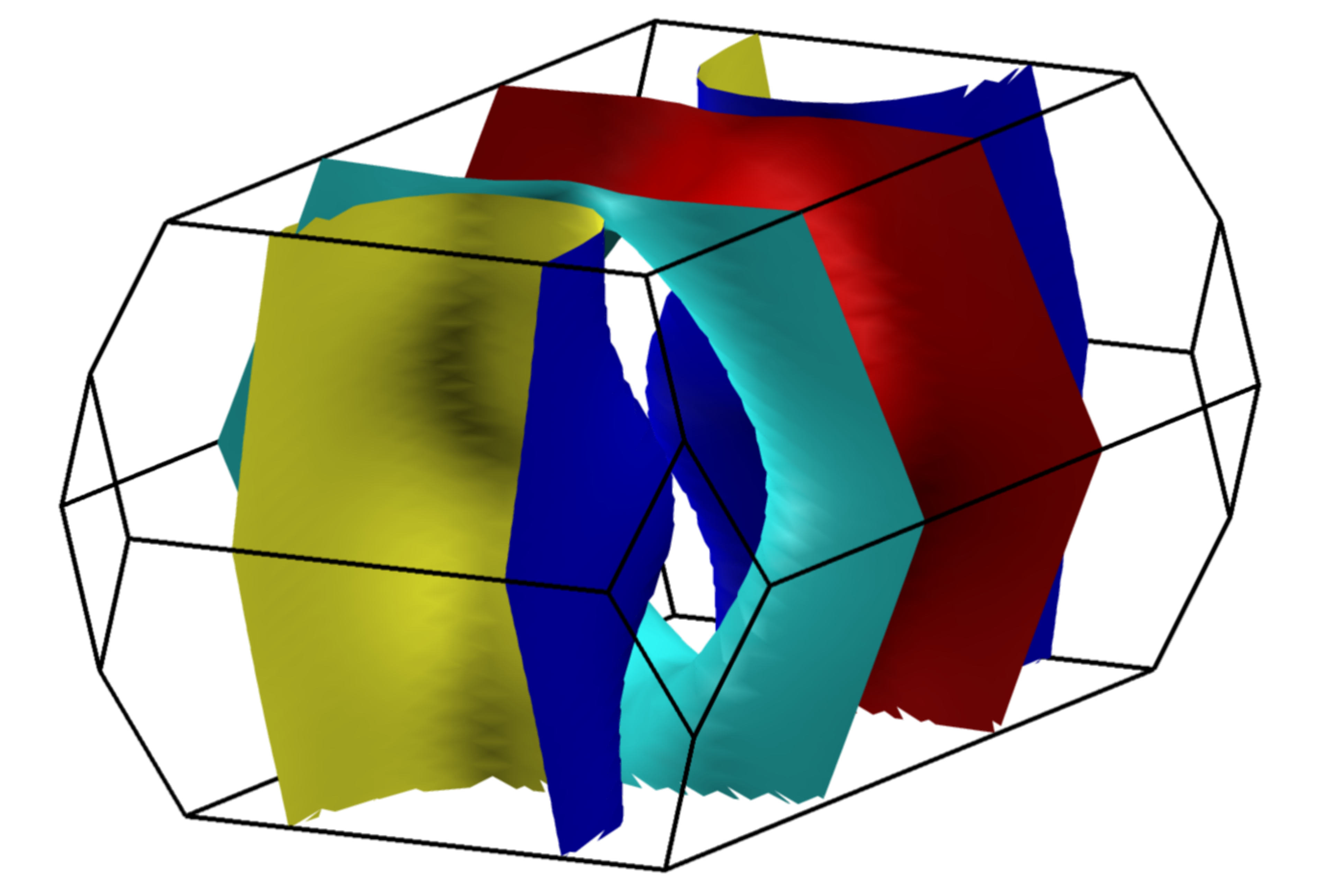}
\centering
\caption{
Fermi surfaces obtained from a DFT$+U$ calculation for $U=1.5$ eV.
\label{fig:FS_DFT_2000k}}
\end{figure}

In Fig. \ref{fig:band} we compare the band structure obtained from the tight-binding Hamiltonian Eq.~(\ref{eq:H}) with that from a DFT$+U$ calculation with an intermediate Coulomb interaction $U=1.5$ eV. 
Although the number of bands is different because we neglect orbital degeneracy in the tight-binding model, flat U 5$f$ band and dispersive U 6$d$ and Te 5$p$ bands are reasonably described. In particular, the low-energy band structure is appropriately reproduced. The FSs (see Fig. \ref{fig:FS_I} in the main text and Fig.~\ref{fig:FS_DFT_2000k}) as well as the weight of U 5$f$, U 6$d$ and Te 5$p$ electrons on the FSs are similar between the tight-binding model and the DFT+$U$ calculation. 


\begin{table*}[tbp]
 \centering
 \caption{Tight-binding parameters for the $f$-$d$-$p$ Hamiltonian.
}
 \label{tab:params}
 \begin{tabular*}{1.0\textwidth}{@{\extracolsep{\fill}}llrlrlrlrlrlr}
   \hline\hline
    & \multicolumn{2}{l}{$f$} & \multicolumn{2}{l}{$d$} & \multicolumn{2}{l}{$p$} &
    \multicolumn{2}{l}{$fd$} &
    \multicolumn{2}{l}{$fp$} &
    \multicolumn{2}{l}{$dp$} \\\hline
    Onsite & $\varepsilon^f_0$ & $0.33$ & $\varepsilon^d_0$ & $0.7$ & $\varepsilon^p_0$ & $-2.2$ & & & & & &\\
    Nearest & $t^f_1$ & $-0.1$ & $t^d_1$ & $-0.6$ & $t^p_1$ & $1.65$ & $t^{fd}_1$ & $-0.05$ & $t^{fp}$ & $0.125$ & $t^{dp}$ & $-0.3$ \\
    2nd Nearest & $t^f_2$ & $-0.075$ & $t^d_2$ & $-0.3$ & & & $t^{fd}_2$ & $-0.1$ & & & & \\
    3rd Nearest & $t^f_3$ & $0.025$ & $t^d_3$ & $0.1$ & & & $t^{fd}_3$ & $-0.025$ & & & & \\
   \hline\hline
 \end{tabular*}
\end{table*}

\section{Magnetic Susceptibility}
\label{ap:chi}
\begin{figure}[tbp]
\includegraphics[width=1.0\linewidth]{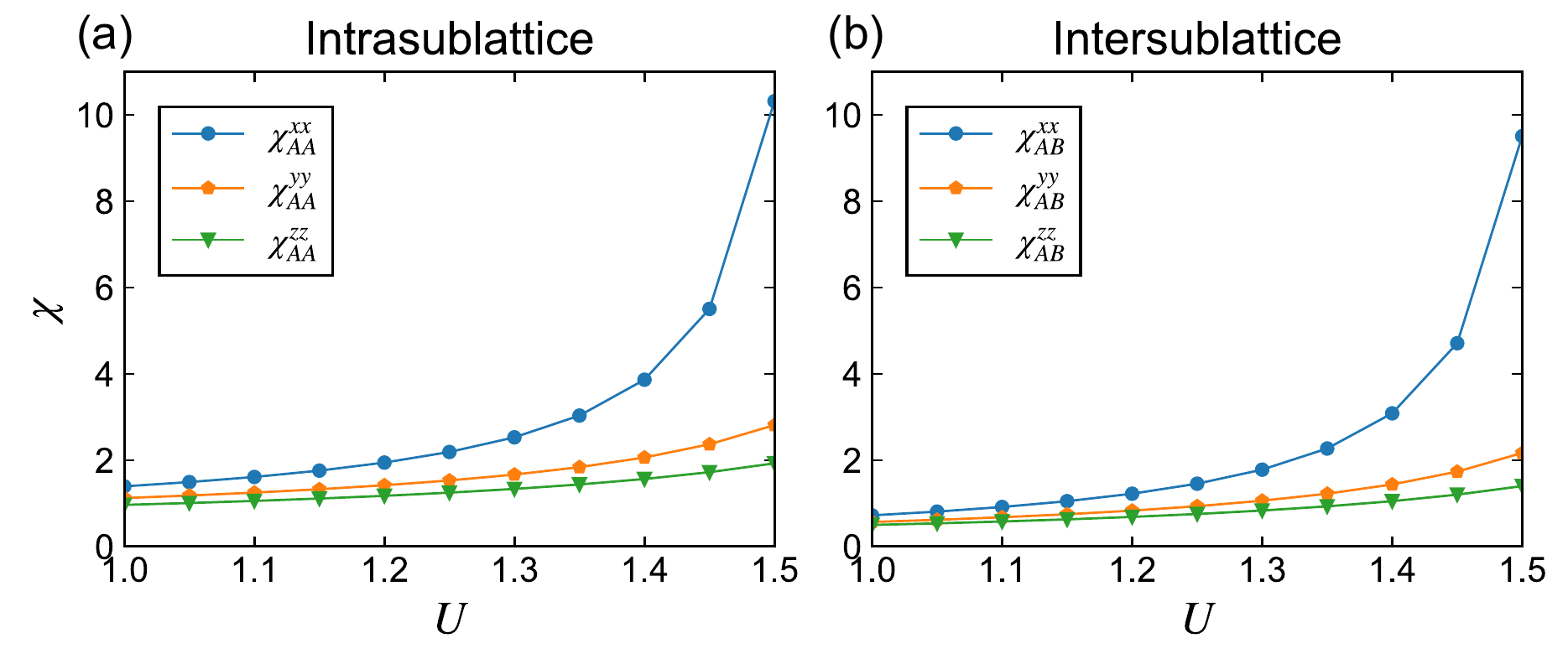}
\centering
\caption{
(a) $U$ dependence of the magnetic susceptibility, $\chi^{xx}_{\rm AA}(\bm 0, 0)$, $\chi^{yy}_{\rm AA}(\bm 0, 0)$, and $\chi^{zz}_{\rm AA}(\bm 0, 0)$ for $p = 1$, $\alpha=0.1$, and $T=0.003$. 
(b) The same plot for the intersublattice susceptibility, $\chi^{xx(yy)(zz)}_{\rm AB}(\bm 0, 0)$.
}
\label{fig:chi_udep_suppl}
\end{figure}

The susceptibility matrix in the $f$-orbital subspace is calculated by the random phase approximation (RPA) as
\begin{align}
    \hat{\chi}(q)=\hat{\chi}^{0}(q)\left[\hat{1}-\hat{\Gamma}^{0} \hat{\chi}^{0}(q)\right]^{-1},
\end{align}
where the irreducible susceptibility is defined as $\hat{\chi}^{0}(q)=$ $-(T / N) \sum_{k} \hat{G}(k+q) \hat{G}(k)$. $\hat G(k)$ and $\hat \Gamma^0$ are the $f$-orbital Green's function and the bare irreducible vertex, respectively.
Here we introduce site-resolved magnetic susceptibilities
\begin{align}
\chi_{m m^{\prime}}^{\mu \nu}(q)=\sum_{s_{1} s_{2} s_{3} s_{4}} \sigma_{s_{1} s_{2}}^{\mu} \chi_{m s_{1} m s_{2}, m^{\prime} s_{3} m^{\prime} s_{4}}(q) \sigma_{s_{4} s_{3}}^{\nu},
\end{align}
for $\mu, \nu = x, y, z$.
We calculate the magnetic susceptibility on Uranium atoms by
\begin{align}
    \chi_{a(b)(c)}(q) = 1/N_{\rm U} \sum_{m} \chi^{xx(yy)(zz)}_{mm}(q),
\end{align}
where $N_{\rm U}$ represents the number of Uranium atoms.
Figure~\ref{fig:chi_udep_suppl} shows the $U$ dependence of the magnetic susceptibility. The magnetic susceptibility and its anisotropy grow with $U$ because the susceptibility is divergent at the magnetic critical point. The larger $U$ (lower $T$) enhances the anisotropy leading to the spin-triplet $B_{3u}$ and $A_u$ superconductivity.


\section{Eliashberg equation and gap function}
\label{ap:sc}

Superconductivity is investigated by solving the linearized Eliashberg equation formulated as
\begin{align}
    \lambda \Delta_{\xi \xi^{\prime}}(k)=&-\frac{T}{N} \sum_{k^{\prime}} \sum_{\xi_{1} \xi_{2} \xi_{3} \xi_{4}} V_{\xi \xi_{1}, \xi_{2} \xi^{\prime}}\left(k-k^{\prime}\right) \nonumber \\
    & \times G_{\xi_{3} \xi_{1}}\left(-k^{\prime}\right) \Delta_{\xi_{3} \xi_{4}}\left(k^{\prime}\right) G_{\xi_{4} \xi_{2}}\left(k^{\prime}\right),
\end{align}
with $\xi = (m, s)$.
The effective pairing interaction is described by the RPA susceptibility as
\begin{align}
    \hat{V}(q) = -\hat{\Gamma}^{0} \hat{\chi}(q) \hat{\Gamma}^{0}-\hat{\Gamma}^{0}.
\end{align}
Solving the Eliashberg equation, we obtain an eigenvalue $\lambda$ and a gap function $\Delta_{\xi \xi^{\prime}}(k)$ for each irreducible representation.
Eigenvalues at a fixed temperature $T=0.003$ are shown in Fig.~\ref{fig:lambda}(a) in the main text, where the Coulomb interaction is set as Table \ref{tab:U} so that sizable eigenvalues are obtained. A similar result is obtained for a fixed $U$. We actually show the $p$-dependence of transition temperatures for a fixed $U$ in Fig.~\ref{fig:lambda}(b). 
The maximum magnitudes of each spin component in the intrasublattice gap function are shown in the main text (Table \ref{tab:gap}), while those of the intersublattice gap function are shown in Table~\ref{tab:gap_inter}.


\begin{table}[htbp]
 \centering
 \caption{Parameters of the Coulomb interaction $U$ used in drawing Fig.~\ref{fig:lambda}(a).}
 \label{tab:U}
 \begin{tabular*}{1.0\linewidth}{@{\extracolsep{\fill}}lcccccc} \hline\hline
  $p$ & $1.0$ & $1.5$ & $2.0$ & $2.5$ & $3.0$ & $3.5$ \\
  $U$ & $1.52$ & $1.92$ & $2.14$ & $2.12$ & $2.05$ & $1.99$ \\
 \hline\hline
 \end{tabular*}
\end{table}

\begin{table}[tbp]
 \centering
 \caption{Maximum magnitudes of intersublattice (A-B sublattice) components of the gap function $d^{\mu}(\bm k)$ obtained from the linearlized Eliashberg equation. The $B_{3u}$, $A_u$, and $A_{g}$ states for $p=1.0$, $2.0$, and $3.0$ are shown. Predominant components are labeled with a star $^\star$.}
 \label{tab:gap_inter}
 \begin{tabular*}{1.0\linewidth}{@{\extracolsep{\fill}}lcccc} \hline\hline
  & $d^{0}$ & $d^{x}$ & $d^{y}$ & $d^{z}$ \\ \hline
  $B_{3u}$ & $\times$ & $1.5 \times 10^{-4}$ & $1.6 \times 10^{-3}$ & $^\star 4.9 \times 10^{-3}$ \\
  $A_{u}$ & $\times$ & $ 1.6 \times 10^{-4}$ & $^\star 1.6 \times 10^{-3}$ & $ 8.7 \times 10^{-4}$ \\
  $A_{g}$ & $^\star 8.9 \times 10^{-4}$ & $\times$ & $\times$ & $\times$  \\
 \hline\hline
 \end{tabular*}
\end{table}


%

\end{document}